\documentclass[english,prl,twocolumn,showpacs]{revtex4-1}
\usepackage[T1]{fontenc}
\usepackage[latin9]{inputenc}
\usepackage{pdfcolmk}
\usepackage{textcomp}
\usepackage{amsmath}
\usepackage{amsthm}
\usepackage{graphicx}
\PassOptionsToPackage{normalem}{ulem}
\usepackage{ulem}

\makeatletter

\usepackage{SIunits}

\@ifundefined{showcaptionsetup}{}{%
 \PassOptionsToPackage{caption=false}{subfig}}
\usepackage{subfig}
\usepackage{babel}

\begin{document}

\title{Spatial instabilities in a cloud of cold atoms}

\author{Rudy Romain}
\altaffiliation{Current address: Department of Physical Sciences, The Open University, Walton Hall, MK7 6AA, Milton Keynes, United Kingdom}

\author{Antoine Jallageas}
\altaffiliation{Current address: Laboratoire Temps-Fr\'equence, Institut de Physique, Universit\'e
de Neuch\^atel, Avenue de Bellevaux 51, 2000 Neuch\^atel, Switzerland}

\author{Philippe Verkerk}

\author{Daniel Hennequin}
\email{Corresponding author: daniel.hennequin@univ-lille1.fr}
\altaffiliation{\\Permanent address: Laboratoire PhLAM B\^at. P5 - Universit\'e Lille, 59655 Villeneuve d'Ascq cedex,  France}
\affiliation{Universit\'e Lille, CNRS, UMR 8523 - PhLAM - Physique des Lasers Atomes et Mol\'ecules, 59000 Lille, France}

\begin{abstract}
Dense cold atomic clouds have been shown to be similar to plasmas. Previous studies showed that such clouds exhibit instabilities induced by long-range interactions. However they did not describe the spatial properties of the dynamics. In this Letter, we study experimentally the spatial nature of stochastic instabilities and find out that the dynamics is localized. Data are analyzed both in the spectral domain and in the spatial domain (principal component analysis). Both methods fail to describe the dynamics in terms of eigenmodes, showing that space and time are not separable. 


\end{abstract}
\maketitle
During the last decades, cold atoms have proven to be more than an
extraordinary tool for studying the physics of dilute matter. Many
spectacular results concern the field of condensed matter, as e.g.
the direct observation of the Anderson localization \cite{jendrzejewski2012},
or that of the BEC-BCS crossover \cite{bourdel2004}. But even in
the field of dilute matter, cold atoms are thought to be a good model
system for plasmas, in particular because experiments are considered
to be relatively simple and well controlled. Although cold atoms in
a Magneto-Optical Trap (MOT) are neutral, it has been demonstrated
that a coulombian-like repulsive force appears in the multiple scattering
regime \cite{pruvost2000}. Based on this analogy, fluid-dynamical models used in plasma physics have been adapted to cold atoms physics \cite{mendonca2008,mendonca2012,rodrigues2016}. On the other hand, it has been demonstrated that the
dynamics of cold atoms in a MOT can be described through the Vlasov-Fokker-Planck equations \cite{romain2011}, as e.g. the plasma dynamics in the inertial confinement fusion \cite{ridgers2008}, the stellar dynamics
\cite{chavanis2003} or the electron dynamics in storage rings \cite{roussel2014}.

Most of these systems are known to exhibit instabilities under appropriate
parameter sets. Numerous types of instabilities have been observed,
with very different properties and signatures. Although plasmas are
governed by long-range interactions, instabilities appear not only
at large spatial scales, but also at smaller scales. Some examples
of local instabilities are the microbunching instability in the storage
rings \cite{roussel2014}, drift wave microinstabilities in plasmas
confined by a magnetic field \cite{weiland2012}, or microinstabilities
of the solar corona \cite{marsch2006}.

Instabilities in MOTs have been observed for several decades, and
have been studied for 15 years. Mainly two types of instabilities
have been reported through experiments. Self-sustained instabilities
are periodic oscillations \cite{labeyrie2006,distefano2003}, while
stochastic instabilities exhibit random characteristics \cite{wilkowski2000}.
In all cases, the experimental characterization is done through the
temporal evolution of global variables, such as the fluorescence of
the cloud or the location of its center of mass. The
spatial properties of the instabilities, in particular their location
in the cloud, is not known. However, most simplified models allowing
to reproduce these instabilities have considered they are global \cite{hennequin2004,distefano2004}.
On the other hand, taking formally into account the different processes
involved in the MOTs leads to the Vlasov-Fokker-Planck equations, implicitly predicting
local instabilities \cite{romain2011}. Models derived from plasmas also predict local phenomena, such as photon bubbles \cite{mendonca2012}, but none of them has been observed yet.
Thus gaining knowledge on the spatio-temporal characteristics of the cloud dynamics appears to be crucial to know which methods and approximations can be used to solve the full set of equations describing the MOT. It could also help to precise the similarities between cold atoms and plasma instabilities.

We report in this paper the experimental observation
of local instabilities, through a novel detection setup allowing to
record the spatio-temporal evolution of the atoms in the MOT. We focus
here on the previously observed stochastic instabilities \cite{wilkowski2000,hennequin2004},
and we show that these instabilities are localized in a limited area
of the cloud. The analysis of the dynamics by two different methods
does not give results consistent with the hypothesis of global temporal
or spatial motion. The paper is organized as follows: after a brief
description of the experimental setup, we analyze the dynamics of
the MOT through global tools, as used in the literature, in order
to clearly establish the type of instabilities we are studying. Then
we analyze the dynamics through two methods: the local temporal analysis
gives information on the motion eigenfrequencies, while the Principal
Component Analysis (PCA) allows to identify spatial eigenmodes.

Our experimental setup is described in detail in
\cite{distefano2003,wilkowski2000,hennequin2004,distefano2004}. The
MOT is a standard Cs MOT with three retro-reflected beams. However,
special care is taken to the stability of parameters which could introduce
artefacts in the dynamics. For example, we use a single mode optical
fiber to clean the transverse profile of laser beams. We also modulate
the relative phases of all the beams to avoid possible interference
patterns. The modulation frequency (> 1 kHz) is chosen larger than
the collective atomic response frequencies, so that the intensity
is averaged \cite{romain2014}. The MOT produces a dense cloud of
cold atoms with a typical diameter of $4\,\milli\meter$. This size
is characteristic of the multiple scattering regime in which the collective
effects play a key role in the behavior of the cloud \cite{sesko1991}.
To obtain the desired dynamical regime, we adjust two control parameters:
(i) the intensity $I$ of a single incoming beam, expressed in units
of the saturation intensity $I_{sat}=1.1\,\milli\watt.\centi\meter{}^{-2}$
(D$_{2}$ line of $^{133}$Cs) and (ii) the detuning $\Delta$ between
the laser frequency and the atomic transition, expressed in units
of the natural line width $\Gamma=2\pi\times5.22\,\mega\hertz$. In
the following, all the illustrations and examples correspond to $I=11\,I_{sat}$
and $\Delta=-1.8\,\Gamma$. They are typical of the dynamics observed
for $10\,I_{sat}\leq I\leq15\,I_{sat}$ and $-1.9\,\Gamma\leq\Delta\leq-1.6\,\Gamma$.

The spatio-temporal dynamics of the atoms is analyzed by recording
the local temporal evolution of the fluorescence at any point of the
cold atom cloud. Let us remember that the number of atoms is proportional
to the fluorescence for a given set of the MOT parameters. As during
an acquisition, the MOT parameters are kept constant, we thus record
the spatio-temporal evolution of the atomic density in the cloud.

It has been shown that instabilities exhibit frequencies ranging from
$1$ to $100\,\hertz$ \cite{distefano2003,wilkowski2000,hennequin2004,distefano2004}.
Thus to record the dynamics, a standard video camera at 30 frames
per second is not adapted. We use a fast video camera (Phantom v7.3
camera from Vision Research) able to record up to 10,000 frames/s.
A set of lenses is used to cover an optical field of $15\times10\,\milli\meter$
and a depth of field of $2.5\,\milli\meter$, well fitting the cloud
size. An important point is to determine if the light recorded by
such a camera comes only from the surface of the cloud, or from any
point inside the cloud. Instabilities grow up when the atomic density
is high enough, i.e. when the collective nonlinear processes inside
the cloud cannot be neglected \cite{distefano2003,wilkowski2000,hennequin2004,distefano2004}.
However, even for such relatively dense clouds, the number of scattering
events for most photons escaping the cloud is 1 or 2 \cite{romain2014}.
That means that the camera captures photons coming from any point
inside the cloud, but with a different weight depending on the location.
This last point has to be kept in mind, although it has marginal consequences
on the interpretation of the pictures. Indeed, the main difficulty
is that we record a 2D projection of the cloud, while the dynamics
occurs in 3D. This makes the interpretation of the records harder.

The location of the camera is imposed by the available space on the setup. Let us call $Ox$, $Oy$ and $Oz$ the three perpendicular axes corresponding to the three incident beams of the trap, $Oz$ being also the axis of the coils. The camera axis is close to the $(x=y,z=0)$ direction. Thanks to the choice of a retro-reflecting beam configuration, and because of the shadow effect \cite{distefano2003,wilkowski2000,hennequin2004,distefano2004}, the amplitude of the dynamics is enhanced in a direction along the line $x=y=2z$. The main direction of the dynamics projects with an angle of $20\text{\textdegree}$ on the picture plane, leading to a satisfactory resolution of the dynamics.

In order to make the link between the present work and the previous
reported observations, we also use a 4-quadrant photodiode (4QP) to
monitor the total cloud fluorescence and the position of the cloud
center of mass, as e.g. in \cite{distefano2003}. The signals from
the 4QP and the pictures from the video camera are recorded synchronously.

As pointed out above, we focus here on the regime of stochastic instabilities.
In \cite{wilkowski2000}, that regime is analyzed through the dynamics
of the number of atoms in the cloud and that of its center of mass.
It is described to be a noisy dynamics cut off with bursts of large
oscillations, and the analysis focused on the main frequency component,
appearing inside the bursts. It is also shown that these instabilities
are induced by a stochastic resonance centered on a given detuning.

In the present work, the unstable regime shows the same characteristics
as in {[}12{]}. In particular, for the parameters given above, instabilities
are maximal for $\Delta=-1.8\Gamma$; the dynamics (number of atoms
and center of mass) is a succession of regular bursts and noisy intervals;
in the bursts, a frequency $\omega_{1}\simeq2\pi\times21\,\hertz$
dominates the dynamics. In order to be more exhaustive than the previous
studies, we also studied the characteristics of the secondary frequency
components. In the burst regime, the second component $\omega_{2}\simeq2\pi\times72\,\hertz$
is two orders of magnitude smaller than $\omega_{1}$. The components
$2\omega_{1}$ and $\omega_{2}-\omega_{1}$ are also present, with
amplitudes similar to that of $\omega_{2}$. In the noisy intervals,
$\omega_{2}$ is present, with the same amplitude as in the bursts.
$\omega_{1}$ is also present, but with an amplitude smaller than
$\omega_{2}$. In summary, $\omega_{1}$ appears mainly inside the
bursts, while $\omega_{2}$ is always present.

For this study, our aim is to characterize the nature -- space dependent
or not -- of the dynamics. Thus we do not need to record long time
series with a good time resolution, as it would be necessary for e.g.
making a topological analysis of the dynamics attractors. We only
need to follow the usual rules of frequency analysis, i.e. to record
at least two points per period. The following results correspond to
$512\,\times\,512$ pixels pictures recorded with a rate of 400 frames$/\second$.
The length of the analyzed sequences is $625\,\milli\second$,
i.e. 250 frames, corresponding to the typical length of the bursts. 

\begin{figure}[t]
\begin{centering}
\subfloat[\label{fig:pic20Hz_2}]{\begin{centering}
\includegraphics[width=4.2cm]{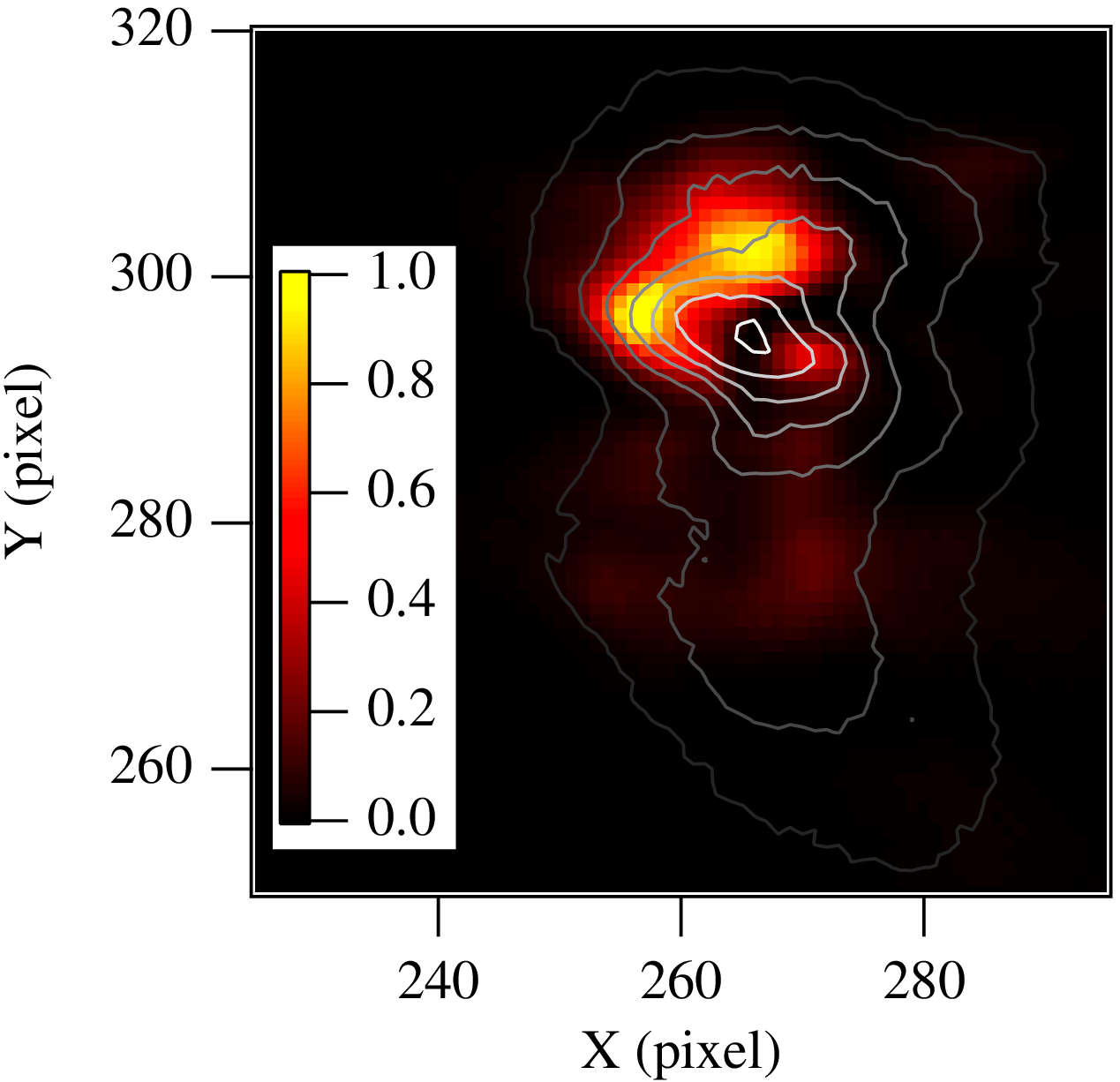}
\par\end{centering}

}\subfloat[\label{fig:pic70Hz_2}]{\begin{centering}
\includegraphics[width=4.2cm]{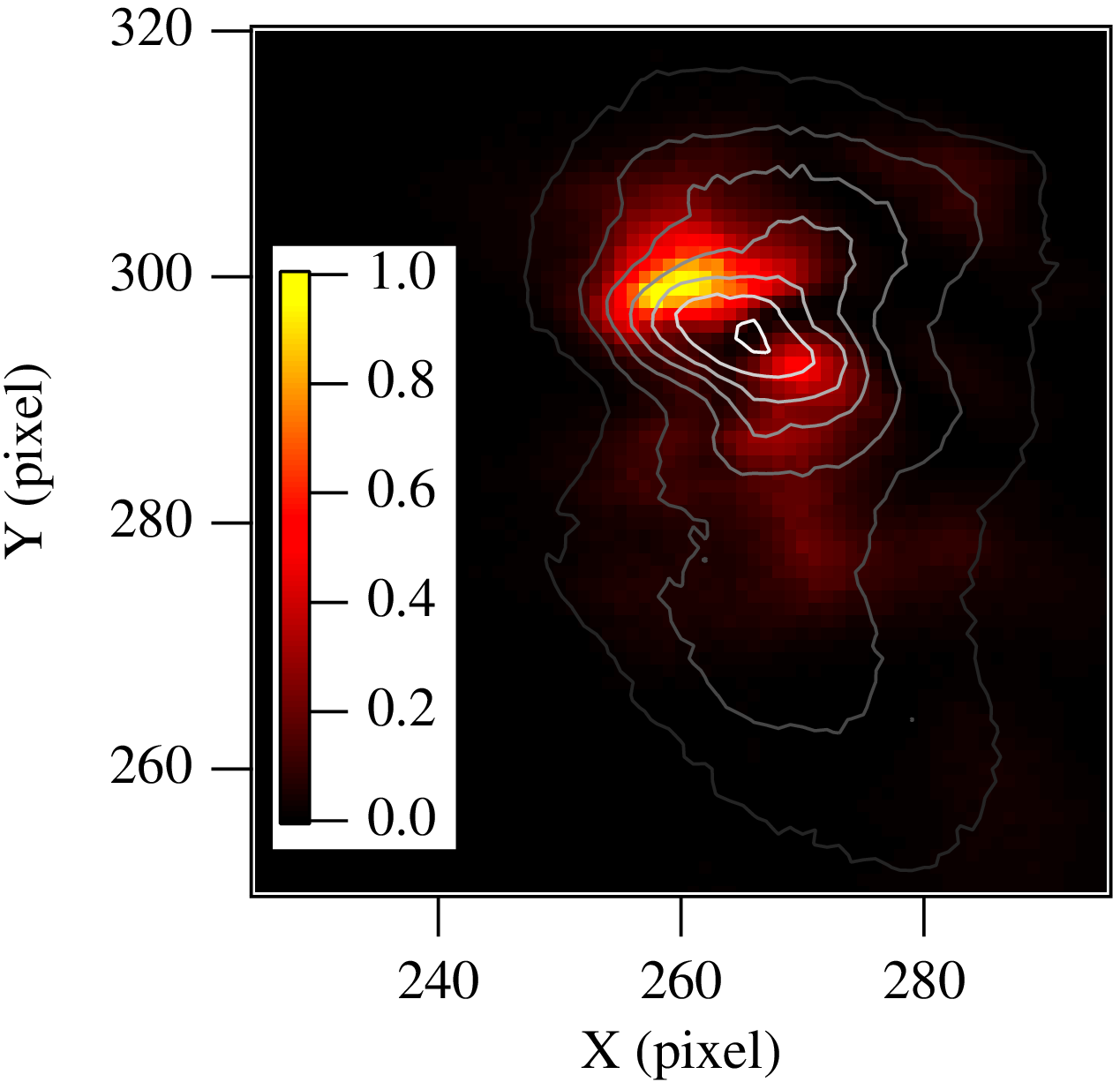}
\par\end{centering}

}
\par\end{centering}

\begin{centering}
\subfloat[\label{fig:phase20Hz_2}]{\begin{centering}
\includegraphics[width=4.2cm]{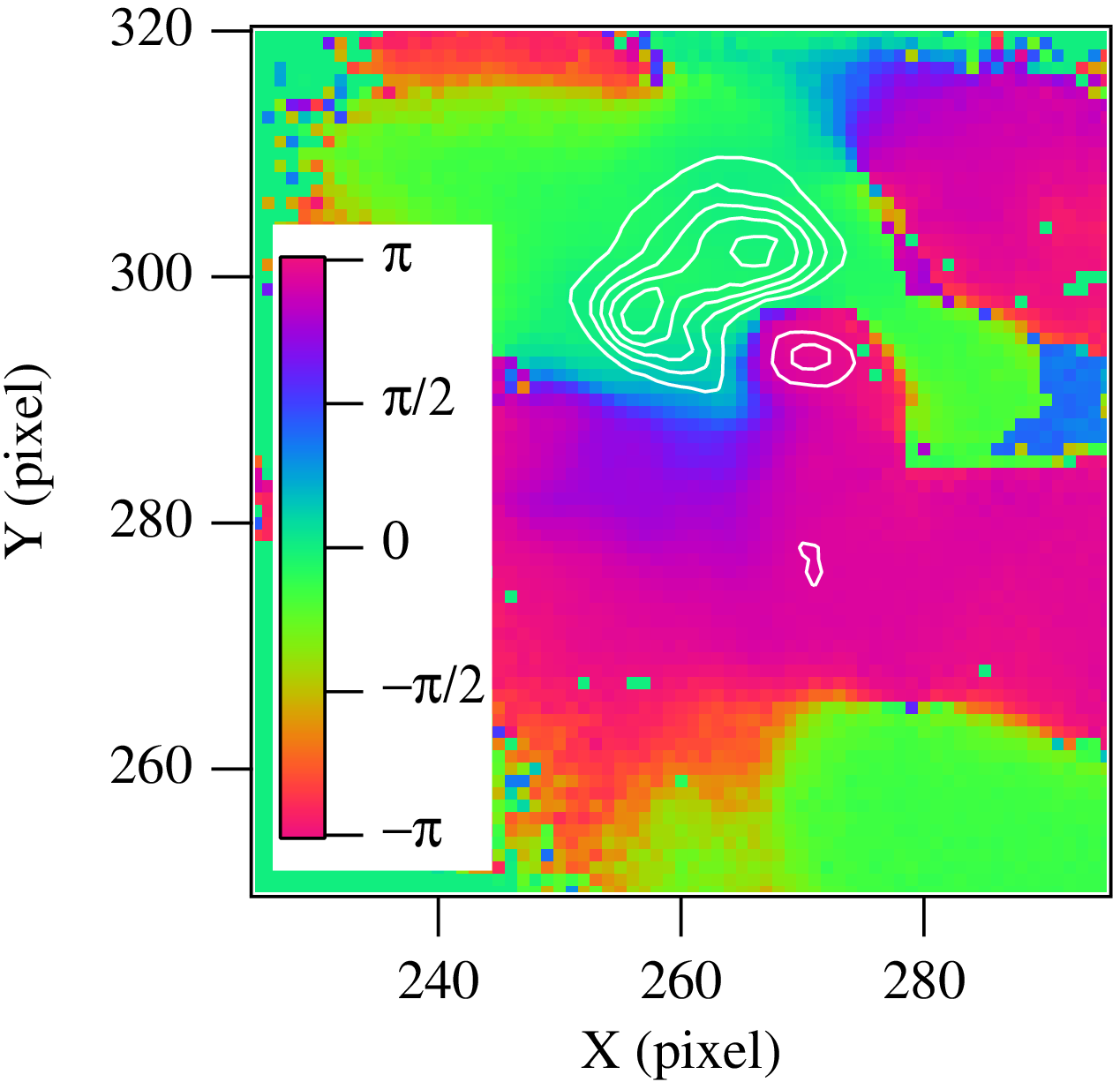}
\par\end{centering}

}\subfloat[\label{fig:phase70Hz_2}]{\begin{centering}
\includegraphics[width=4.2cm]{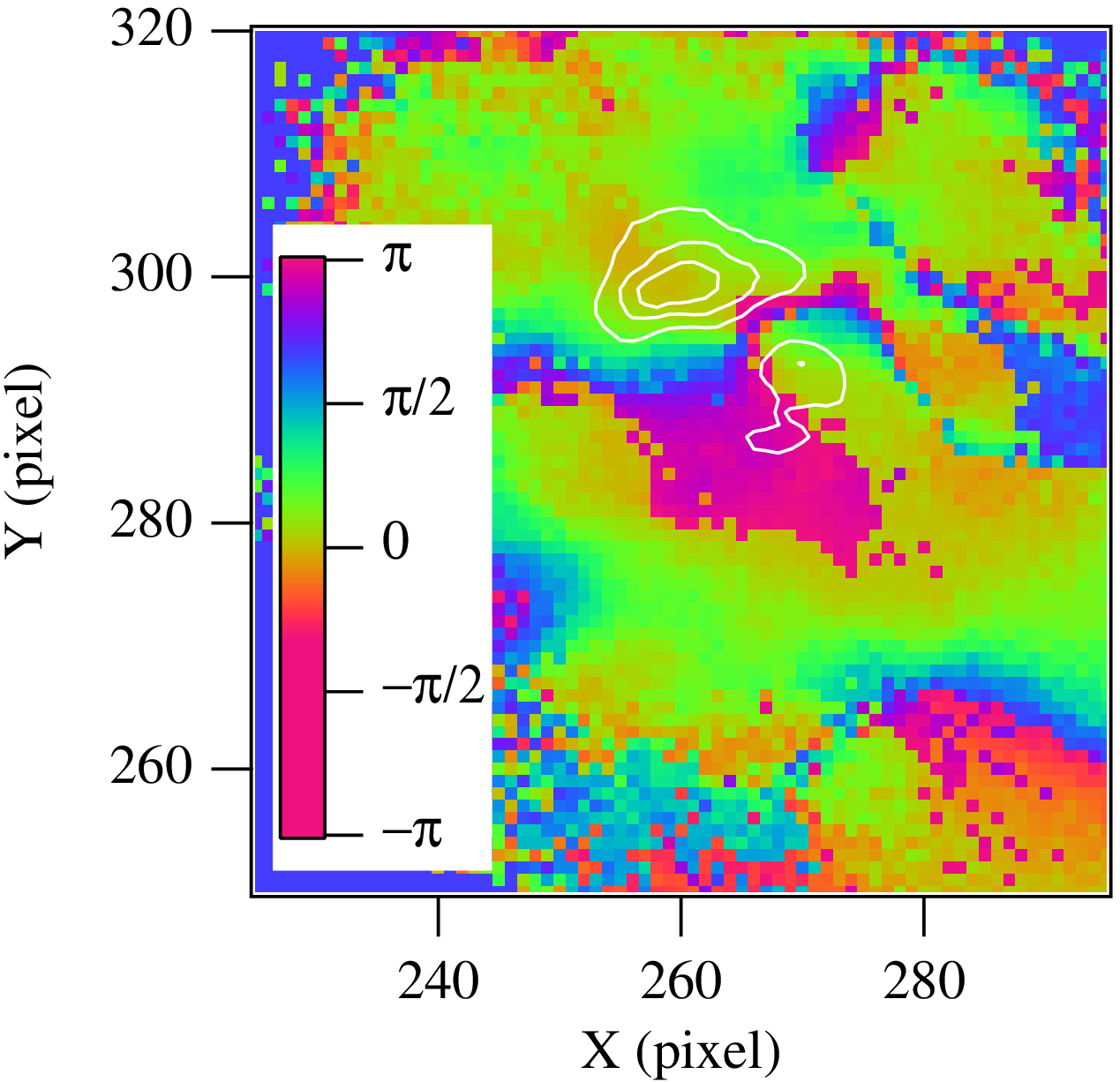}
\par\end{centering}

}
\par\end{centering}

\caption{
Spatial distribution of instabilities for one sequence: (a) and (b)
show the local amplitude of the two main components: $\omega_{1}$
and $\omega_{2}$. The normalized magnitude squared of the spectral
component is represented in gray level (Color online). The contour
plot shows the averaged fluorescence during the sequence. (c) and
(d) represent the corresponding phase distributions. Grayscale (Color
online) represents the relative phase of the local oscillation between
$\pm\pi$. The contour plot identifies the unstable areas.\label{fig:maps}
}
\end{figure}

To determine the spatial dependence of the dynamics, a straightforward
approach consists in applying the frequency analysis used previously to
each point of the cloud. Fig. \ref{fig:maps} shows a typical result
for the dynamics inside a burst. In fig. \ref{fig:pic20Hz_2}, we
represent in gray levels (Color online) the amplitude of the $\omega_{1}$
component in each point of the cloud, obtained by Fast Fourier Transform
(FFT) of each pixel of the picture sequence. The dynamics appears
with no doubt to depend on the space. Instabilities appear in two
small areas of the cloud, covering typically $10\,\%$ of the whole
cloud. The two areas have very different shapes and sizes, and we
see two local maxima in one of the area. Fig. \ref{fig:phase20Hz_2}
shows that the two areas pulse with opposite phases. It is difficult
to deduce from such 2D observations the exact 3D dynamics, but it
is clear that the instabilities consist in a local pulsation or rotation,
while the rest of the cloud is stable.

The above example is typical of what we observed in all recorded sequences.
Some differences appear, as the location of the unstable area, its
structure, in particular the number of local maxima, and its shape,
but we always observed that the instabilities were localized in a
limited area of the cloud, covering typically $10\,\%$ of the whole
cloud.

We performed the same analysis for the $\omega_{2}$ component, and
found the same type of results as for the $\omega_{1}$ component:
the instabilities are localized in a relatively small area, and they
correspond to a motion of pulsation or rotation of a part of the cloud.
As discussed above, inside the bursts, the $\omega_{2}$ component
is small as compared to the $\omega_{1}$ component, making the analysis
of data less reliable. However, it appears clearly that in this case,
the characteristics of the $\omega_{2}$ component follow those of
the $\omega_{1}$ component, delimiting the same unstable area with
the same type of motion. Figs \ref{fig:pic70Hz_2} and \ref{fig:phase70Hz_2}
show respectively the amplitude and the phase of the $\omega_{2}$
component for the same burst as figs \ref{fig:pic20Hz_2} and \ref{fig:phase20Hz_2}:
the spatial distribution of the instabilities are the same for the
two components. The similarity of the phase distributions is less
convincing, due to the weakness of the component, but in spite of
that, a phase opposition appears between the two areas. To generalize
this observation, we computed the spatial overlap between the two
components for all the recorded sequences, and found that they have
in most cases a similar spatial distribution.

Outside the bursts, the $\omega_{2}$ component is still present,
and has the same characteristics. Thus the dynamics appears to be
a small (in space) and weak (in amplitude) periodic motion of a small
part of the cloud at the $\omega_{2}$ frequency, cut off by bursts
where the amplitude of the motion increases temporarily, while its
main frequency shifts to $\omega_{1}$. However, this description
is not completely satisfactory. Indeed, it gives information about
the temporal components of the dynamics, but does not give any information
about the spatial components. In particular, it seems to associate
one spatial component with two different frequencies, while in spatio-temporal
systems where the temporal dynamics and the space distribution can
be separated, as e.g. in multimode lasers \cite{hennequin1990}, a
spatial eigenmode is associated with only one eigenfrequency. We adopt
in the following another approach.

PCA presents the advantage to give a description of the dynamics of
the system without requiring any preliminary hypothesis. The dynamics
is described in terms of a superposition of spatial modes, giving
complementary information with respect to the Fourier analysis. The
analysis is performed using the method described in \cite{dubessy2015}.

The result of the PCA is a set of spatial modes forming a basis whose
size is equal to the number of pictures of the sequence (250
in our case). If these modes
are sorted according to their weight, i.e. their contribution to the
total statistical deviation around the averaged atomic distribution,
the method leads to the determination of the number of modes useful
to describe the dynamics. In our case, we find that in the sequences
considered here, the main mode contains between $50\,\%$ and $80\,\%$
of the statistical fluctuations, and the second mode between $5\,\%$
and $25\,\%$. Thus the dynamics appears to be dominated by one single
mode, although the weight of the second mode is not negligible. Depending
on the desired accuracy, the third mode, with a weight of $10\,\%\pm5\,\%$,
could be taken into account. In the following, we consider only the first two modes. 

\begin{figure}[t]
\centering{}\subfloat[\label{fig:first_mode}]{\begin{centering}
\includegraphics[width=4.2cm]{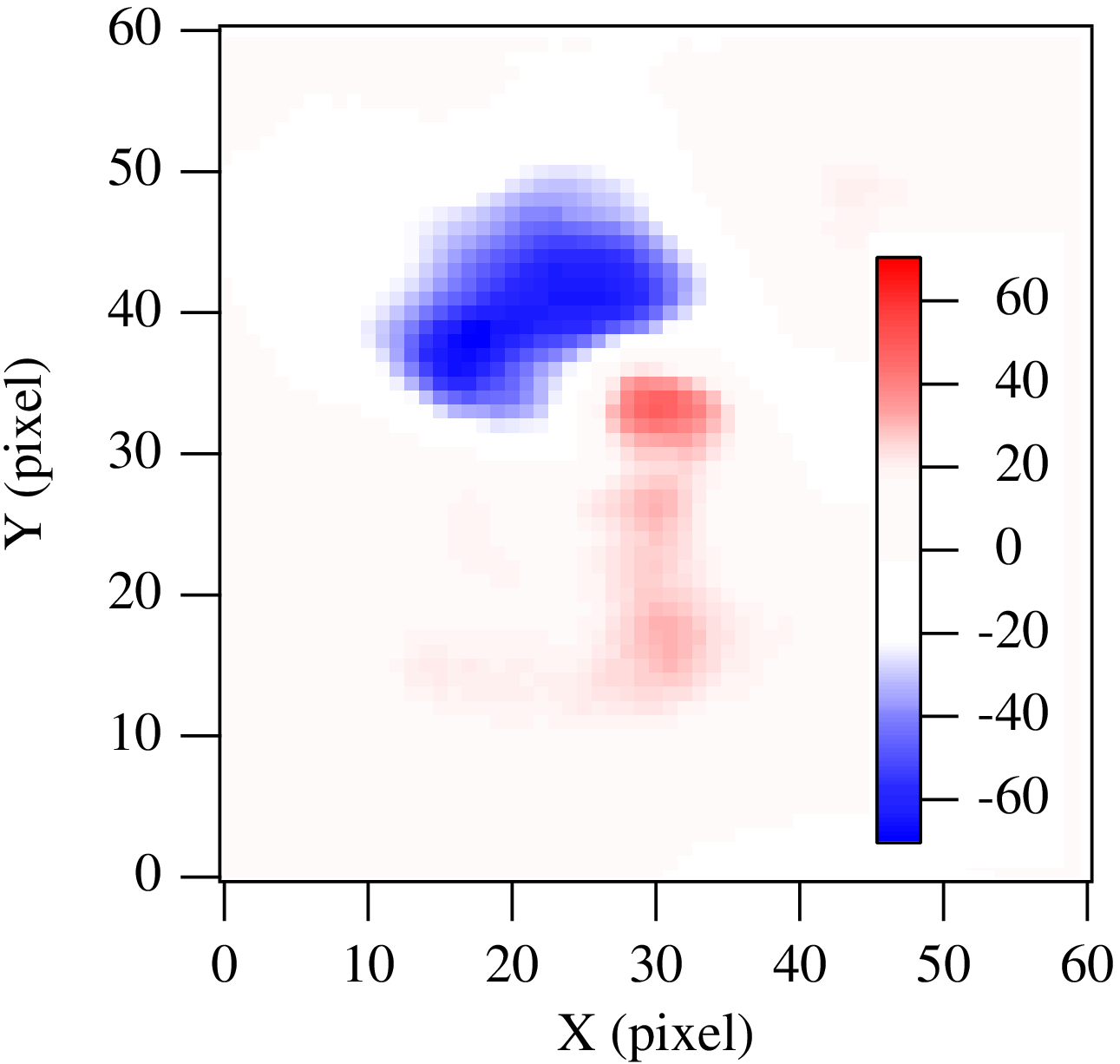}
\par\end{centering}

}\subfloat[\label{fig:second_mode}]{\begin{centering}
\includegraphics[width=4.2cm]{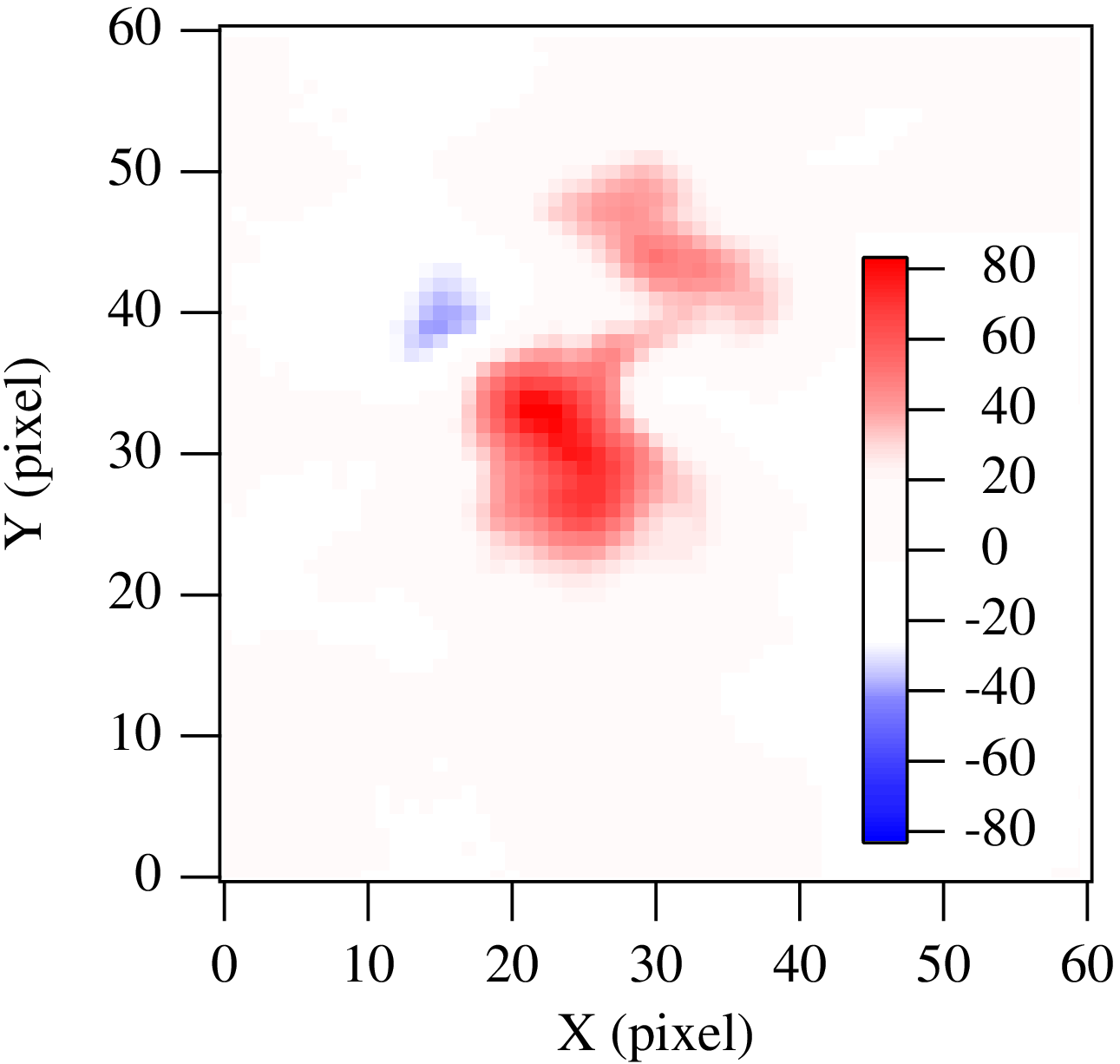}
\par\end{centering}

}\caption{Two first spatial modes given by the PCA (\ref{fig:first_mode} and
\ref{fig:second_mode} respectively) for the same sequence than in
figure \ref{fig:maps}. In both cases, the spatial distribution is
represented in absolute value in gray level (Color online), and the
upper and the lower areas have to be understood respectively as an
excess (in blue online) and a depletion of atoms (in red online).
This information is similar than the one given by the phase of the
Fourier analysis.\label{fig:modesPCA}}
\end{figure}

Figure \ref{fig:modesPCA} shows the first two modes obtained by the
PCA of the sequence of figure \ref{fig:maps}. Note the similarity
of the main mode with the $\omega_{1}$ spatial distribution (fig.
\ref{fig:pic20Hz_2}). However, this result is not general. On the
contrary, we observe a major difference between the modes given by
the PCA and the ones resulting from the Fourier analysis. While in
the latter a wide variety of shapes are obtained, the results are
more homogeneous in the former. In fact, one set of modes dominates
the whole dynamics. This basis -- let us call it the main basis --
is not the only one found by the PCA, but it is the most frequent
in the set of recorded sequences. This regime is randomly interrupted
by short intervals where the dynamics is described by another basis,
before it returns to the main basis. These alternative bases differ
from one sequence to another one, and we did not found any common
properties. For example, in a few cases, the modes of the main basis
appear, but in a different order: the first mode of the main basis
becomes the second mode of the basis.

\begin{figure}[t]
\begin{centering}
\subfloat[]{\begin{centering}
\includegraphics[width=4.2cm]{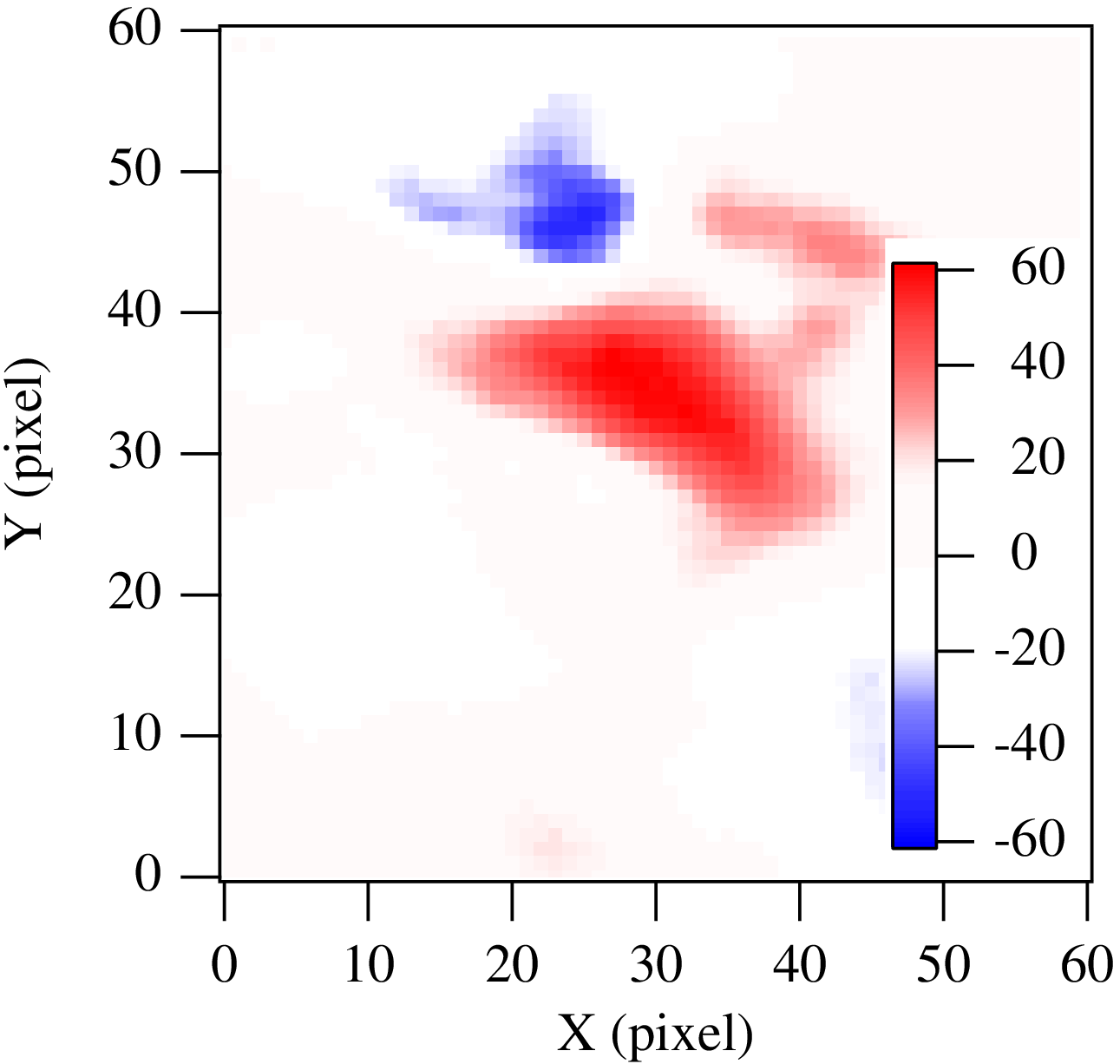}
\par\end{centering}

}\subfloat[]{\begin{centering}
\includegraphics[width=4.2cm]{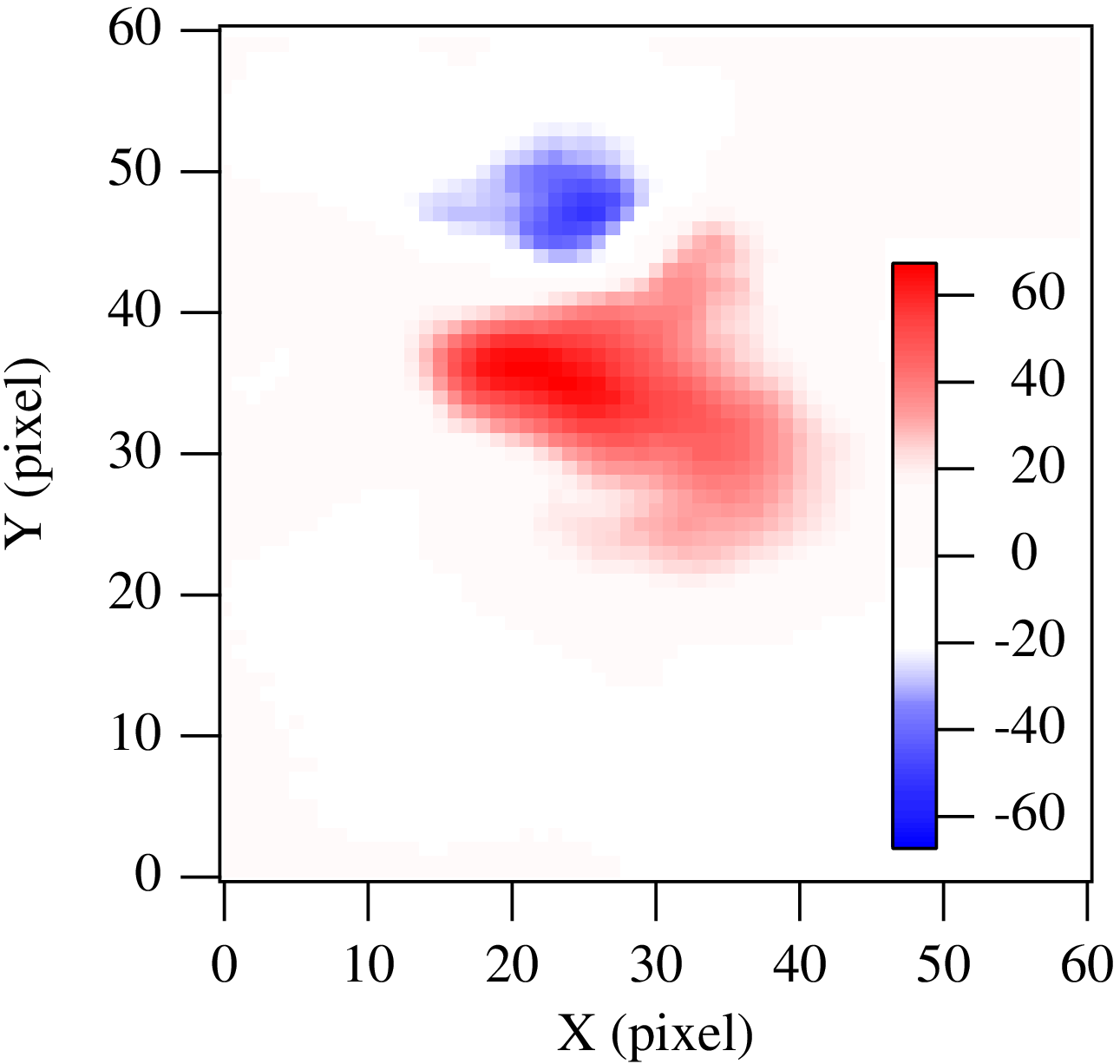}
\par\end{centering}

}
\par\end{centering}

\begin{centering}
\subfloat[]{\begin{centering}
\includegraphics[width=4.2cm]{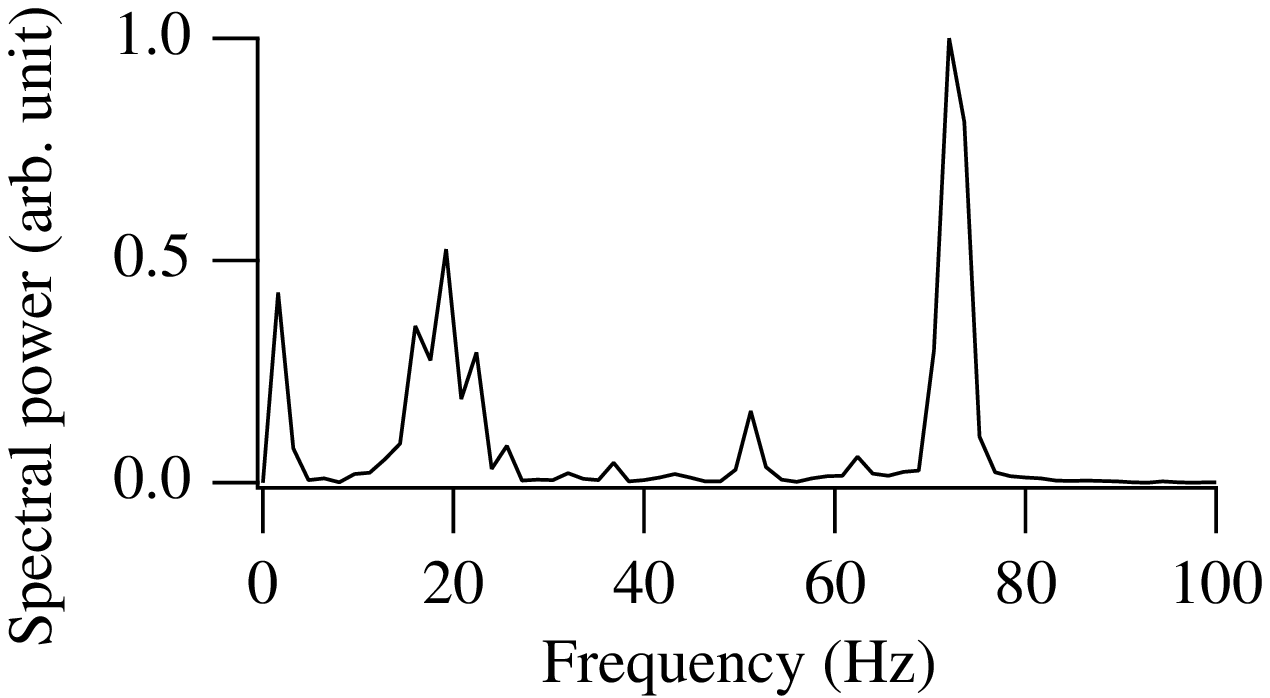}
\par\end{centering}

}\subfloat[]{\begin{centering}
\includegraphics[width=4.2cm]{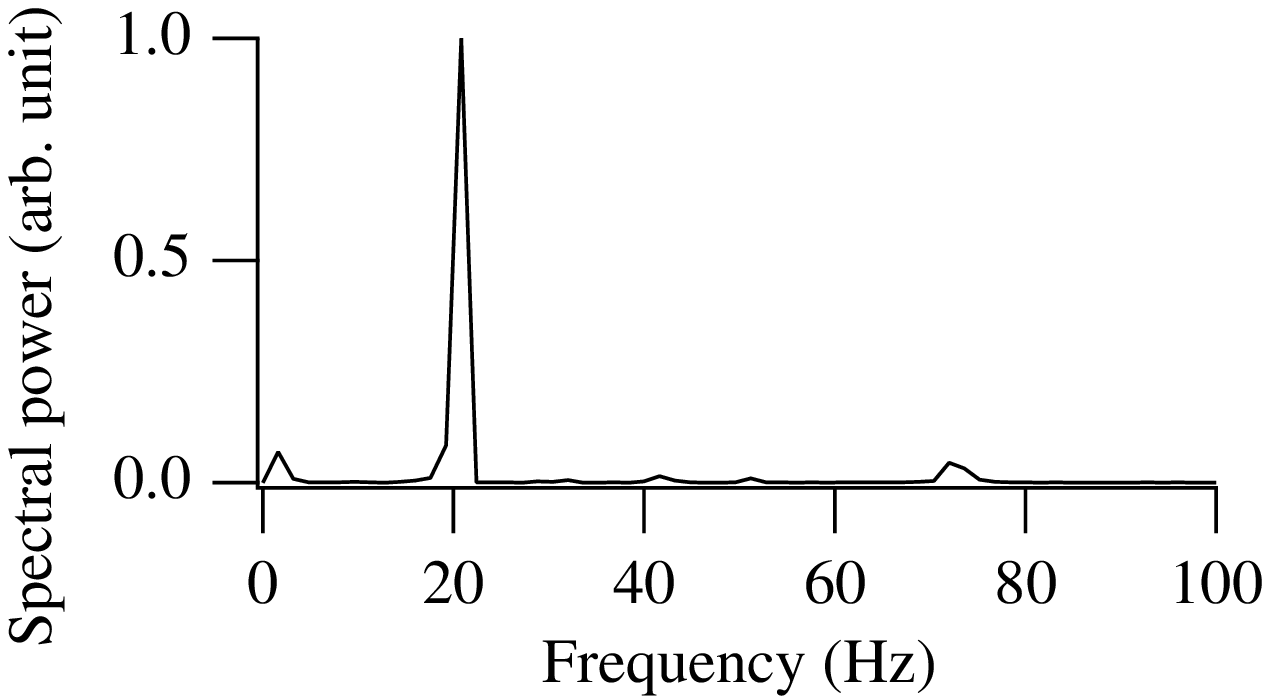}
\par\end{centering}

}
\par\end{centering}

\centering{}\caption{From top to bottom: first modes calculated by the PCA for two consecutive
sequences and the FFT of the time evolution of these modes.\label{fig:TF mode}}
\end{figure}

A surprising result is that the regimes found with the PCA do not
correspond to those given by the Fourier analysis. In particular,
the sequences described by the main basis do not follow the succession
of bursts and noisy intervals. As a consequence, the frequencies associated
with the main basis may change from one sequence to another. Moreover,
the time evolution of a given spatial mode may exhibit different frequencies
in different sequences. Fig. \ref{fig:TF mode} shows the time evolution
and the FFT of two modes dominating the dynamics in two consecutive
sequences. Although the two modes are similar, they are associated
with two very different time evolution: for the first one, both frequencies
$\omega_{1}$ and $\omega_{2}$ drive the dynamics, while the second
one corresponds to a burst where $\omega_{1}$ dominates the dynamics.

In summary, the PCA leads to a description of the dynamics in terms of spatial
modes. The results show that such a description is as relevant as
the description in terms of frequency components. Two regimes are
found: the first one corresponds to a well identified spatial mode
basis, referred as the main basis, and a second one, described by
different bases. From this point of view, the description obtained
by the PCA is similar to that of the Fourier analysis. But surprisingly,
the two regimes of the PCA do not correspond in time sequences to
the two regimes of the Fourier analysis. Thus the two approaches --
temporal and spatial -- give complementary information on the dynamics:
the dynamics of the cloud of cold atoms in a MOT is a genuine spatio-temporal
system, where the spatial and temporal behaviors cannot be separated.

We report in the present paper experimental results on the dynamics of an unstable cloud of cold atoms in a regime of stochastic instabilities. Previous studies focused on the temporal behavior of the instabilities. Here we study both the spatial and temporal properties of the dynamics. Although the atomic motion cannot be clearly identified as our analysis is based on a 2D projection of a 3D motion, we show that the oscillations are localized in space. We analyze the dynamics through two different methods, and both point out the key role of space in the dynamics. Moreover, the analyses in terms of frequency components and in terms of spatial modes show that the relation between the temporal regime and the spatial distribution is not straightforward, as the same spatial distribution can correspond to different temporal regimes. These results invalidate the description in terms of purely temporal models, as in \cite{labeyrie2006,distefano2003,wilkowski2000,hennequin2004,distefano2004}, and require the use of fully spatio-temporal models, as the Vlasov-Fokker-Planck model  \cite{romain2011} or plasma-derived model \cite{mendonca2012}. They strengthen the relationship between cold atoms and plasmas, and show that cold atoms could be a good model system for plasmas. However, a fine comparison of both systems and a deep physical interpretation of the observed behavior require a better knowledge of the dynamics of cold atoms. So, the next step is to run numerical simulations to obtain quantitative agreement between models and experimental observations, in particular in the dynamical regimes.

\section*{Acknowledgments}

The authors would like to thank R. Dubessy for helpful discussions
about the PCA. This work was supported by the Labex CEMPI (Grant No. ANR-11-LABX-0007-01) and ``Fonds Europ{\'e}en de D{\'e}veloppement Economique R{\'e}gional''.

\end{document}